# Service Health Engineering for Distributed Systems

Siva Rama Krishna Varma Bayyavarapu
Docusign Inc.

**THE INTRODUCTION** Distributed systems support many critical business workflows, but service health is often judged through component dashboards rather than through a reliability practice focused on end-to-end user outcomes. In complex systems, individual services and nodes can appear healthy while user-facing workflows quietly fail, stall, or degrade. A more useful approach is to treat service health engineering as a practical reliability discipline that connects telemetry, workflow completion, dependency behavior, operational readiness, and recovery validation into a repeatable practice for distributed systems.

To keep the discussion concrete, we use a document approval workflow as a running example. In a typical organization, documents are submitted, routed through configured approvers, and expected to reach a clear final state like approved, rejected, or cancelled, within an acceptable time window. That journey may traverse user interfaces, microservices, queues, background jobs, databases, notification systems, and support processes. The question is not only whether each component is "up," but whether the service keeps its promise to the user who submitted the document.

Site reliability engineering (SRE) has shown how production operations can be managed as an engineering discipline rather than through ad hoc firefighting [1], [2]. Service Health Engineering does not replace service-level objectives (SLOs), operational reviews, or chaos engineering. It brings these established practices together around end-to-end journey completion, particularly in asynchronous systems where work can stall without an obvious component-level failure. The approach defines service promises at the journey level, embeds reliability criteria into design and delivery, uses diagnostic measures such as Mean Time to Detect (MTTD), Mean Time to Recover (MTTR), and customer impact time, and establishes recurring practices such as weekly service-health reviews and resiliency validation. It also describes how a human-reviewed, artificial intelligence (AI)-assisted reporting architecture can help teams assemble health evidence without turning AI into an autonomous decision-maker.

## SERVICE PROMISES AND CUSTOMER JOURNEYS

Service health, in this context, is about whether a service keeps its promise to the user. For the document approval workflow, the service promise might be stated as: Submitted documents are routed through configured approvers and reach a clear final decision within an expected time window, with transparent status visible to the requester.

A service promise states the outcome in user terms. Service-level indicators (SLIs) measure that outcome, and SLOs define the target. For the approval journey, completion rate and time-to-decision are SLIs; the SLO specifies the acceptable completion percentage and latency bound.

This promise lives at the journey level, not at the level of individual components. It focuses on the outcome that matters to users: whether their document progresses through the approval process, how long that process takes, whether the final decision is correct, and whether status is visible and accurate along the way.

In large enterprise Software-as-a-Service (SaaS) environments, these journeys often span multiple teams, services, and regions. An approval can depend on tenant-specific configuration, regulatory routing rules, and external notification channels. The promise still needs to be stated in terms users understand: the document will reach a decision within an agreed service window, and the requester can track its status, even if the underlying path differs by tenant, region, or risk profile. When reliability discussions stay at the level of CPU charts or generic error rates, the gap between internal views and user expectations grows.

Journey-level signals complement rather than replace component telemetry. CPU utilization, API error rates, and queue depth help explain system behavior, while completion rate and time-to-decision show whether the approval workflow delivered the promised outcome. Individual services can appear healthy even when approvals remain stalled in background processing or because of a downstream dependency failure.

## WHY COMPONENT-LEVEL MONITORING FALLS SHORT

Modern observability stacks provide an impressive amount of telemetry: metrics, logs, traces, dashboards, and alerts across infrastructure and application components. However, component-level monitoring can miss important workflow-level failures in the document approval journey.

A common pattern is partial failure. The submission API remains available, and the status-query endpoint responds successfully. Dashboards for the approval routing service and the database show normal error rates and latencies. Yet documents are stuck in an intermediate state because a background job that advances approvals is misconfigured or stalled. Component-level monitoring indicates health, but the end-to-end workflow is broken.

Another pattern involves asynchronous workflow gaps. Approvals often rely on queues and scheduled jobs that move documents between states, trigger reminders, or send notifications. Metrics may show that queues are processing messages and that consumers are running, but those metrics do not guarantee that each document traverses all required states. A subset of approvals can become stranded in "pending" or "in review" without any component crossing an alert threshold.

Threshold-heavy alerting can also obscure customer impact. To reduce noise, teams often set high thresholds on error rates, latency, or saturation. Rare but high-impact failures—for example, approvals in a particular region that consistently take many hours—may not cross those thresholds. Operators are paged only when the system is clearly "on fire," leaving moderate but sustained promise violations undetected.

Finally, many systems lack customer impact context in their telemetry. Metrics and logs count events, but do not track how many user journeys fail, how long failures last, or which segments are affected. Ownership follows components: one team owns the submission service, another owns the notification service, another owns the database. No one explicitly owns the reliability of the document approval journey end-to-end.

Service health engineering addresses these gaps by connecting component telemetry to workflow completion, customer impact, and operational readiness. Monitoring becomes one input among many in a broader reliability practice focused on keeping promises to users.

## FIVE DIMENSIONS OF SERVICE HEALTH

To make service health engineering practical, it helps to break the problem into a small number of dimensions that can be designed, instrumented, and reviewed systematically. In production SaaS and enterprise software systems, service health can be viewed through five practical lenses:

- User journey and experience.
- API and service contract behavior.
- Back-end workflow completion.
- Dependency and infrastructure behavior.
- Operational readiness and recovery paths.

The document approval workflow can be examined through each of these lenses.

## User Journey and Experience

At the journey level, the key questions are: Do submitted documents reach a clear decision? How long does it take? What do users see when something goes wrong?

Journey-centric signals include approval completion rate, time-to-decision distributions, and the frequency of stalled approvals or unclear statuses. If a significant fraction of documents remain in intermediate states beyond expected time windows, service health is degraded even if all components report acceptable metrics.

## API and Service Contract Behavior

APIs and service contracts are the interfaces through which workflows are initiated, advanced, and queried. For the approval service, this includes submission endpoints, approval-action APIs, and status-query endpoints.

Reliability-relevant signals at this level include submission and status-query availability, latency percentiles for critical APIs, and error-rate breakdowns that distinguish client errors from server or dependency failures. Contract behavior also matters: idempotent operations simplify retries; clear error payloads help callers understand whether a failure is transient or permanent; and trace propagation across services allows end-to-end correlation of approval requests and decisions [3], [5].

## Back-End Workflow Completion

The approval journey depends heavily on back-end processes: queues that carry approval tasks, background workers that update documents from "pending" to "in review" to "approved" or "rejected," scheduled jobs that trigger reminders, and reconciliation jobs that check for inconsistencies.

Signals here answer whether internal tasks progress through all required states. Signals include the distribution of approvals across states and the number and age of documents in long-lived intermediate states. A workflow watchdog is an independent, scheduled check that queries workflow state and flags items that remain in an intermediate state beyond a journey-specific threshold. A healthy component stack can still be unreliable if approvals are routinely trapped between states without detection.

## Dependency and Infrastructure Behavior

Approval services sit atop databases, caches, messaging systems, notification providers, and compute infrastructure. Traditional health metrics such as availability, error counts, latency, and saturation remain important. Service health engineering adds the question "How does dependency behavior affect approval promises?"

Teams look at dependency-specific signals tied to the approval journey: latency for critical database queries, success rates for notification deliveries, queue depth and lag for approval tasks, and resource saturation across the stack. Behavior under degraded conditions, including failover and fallback strategies, is also part of the picture [4], [5].

## Operational Readiness and Recovery Paths

Operational readiness describes the ability to detect, diagnose, mitigate, and recover from failures that affect approvals. Recovery paths are the engineered mechanisms for restoring service promises when they are violated.

Signals include alert and watchdog coverage, the time required to detect new failure modes, evidence that runbooks are maintained and used, and evidence that rollback, failover, and reprocessing procedures have been tested. A service is not reliably healthy if incidents are difficult to detect, diagnose, or recover from [1], [2], [10].

## SERVICE-HEALTH DIMENSIONS AND EXAMPLE SIGNALS

**Table 1** summarizes these dimensions, using the document approval workflow to illustrate example signals and the reliability questions they answer.

The same dimensions apply to other asynchronous workflows. For payments, the journey runs from authorization through settlement, with duplicate or stranded transactions treated as failures. For provisioning, success means that the requested resource becomes available or the request is rolled back cleanly. For a data pipeline, the expected datasets must arrive on time, in full, and pass validation. The signals vary by workflow, but the service-health dimensions remain the same.

Table 1. Service-Health Dimensions and Example Reliability Signals

| Dimension | Example Signals | Reliability Question Answered |
|---|---|---|
| User journey and experience | Approval completion rate; time-to-decision percentiles; rate of stalled approvals | Are users' approval requests reaching clear decisions within expected time? |
| API and service contract behavior | Submission API availability; status query latency; error-rate breakdown (client/server); idempotent retries | Are callers able to submit and query approvals reliably and predictably? |
| Back-end workflow completion | Count of approvals in each state; watchdog alerts for long-lived "pending"; workflow completion latency | Are internal approval tasks progressing through all required states? |
| Dependency and infrastructure | Database query latency; notification success rate; queue depth and lag; resource saturation | Are critical dependencies and infrastructure supporting the approval promise? |
| Operational readiness and recovery | Alert coverage; time from detection to mitigation; runbook usage; rollback and reprocessing evidence | Can the team detect, mitigate, and recover when approvals are impacted? |

## RELIABILITY CRITERIA IN DESIGN AND DELIVERY

Reliability is often mentioned in nonfunctional requirements but left out of concrete acceptance criteria. Service health engineering brings reliability directly into design and delivery, particularly for critical workflows such as document approval.

Consider the following design-time criteria:

**Trace propagation** ensures that the approval journey carries a trace context across submission APIs, queues, background jobs, notifications, and status queries. This allows engineers to observe end-to-end journeys during incidents and weekly reviews, rather than inferring behavior from isolated logs [3], [9].

**Watchdog signals** provide explicit checks that regularly scan for approvals stranded in intermediate states beyond agreed thresholds. These signals are essential for catching silent failures in asynchronous workflows where no single component metric indicates a problem.

**Bounded retries** and idempotency govern how interactions with dependencies such as notification providers and external approval steps behave under stress. Retries should use sensible limits and backoff, and critical operations such as document submission and final decision updates should be idempotent. This reduces the risk of duplicate approvals, stuck workflows, or cascading failures [5], [7].

**User-facing failure** behavior defines what users see when the approval promise cannot be kept. Clear error messages, accurate statuses (for example, "approval delayed due to system issue"), and transparent communication about expected recovery are part of reliability, not afterthoughts.

**Support traceability** ensures that support and operations teams can link tickets and user reports to specific approval workflows, trace IDs, and underlying failures. Without such traceability, teams spend more of the detection and recovery interval on manual investigation.

When acceptance criteria include reliability behaviors alongside functional outcomes, teams can test reliability during development and pre-production. Synthetic approvals can exercise trace propagation, watchdog checks, retry behavior, and recovery paths before real users encounter failures.

Journey-level telemetry has an operational cost. Workflow, tenant, and trace identifiers can create high-cardinality data, while full-fidelity traces and long retention increase storage and query expense. Instrument the journeys and state transitions tied to an SLO or operational decision. Use metrics for aggregate counts and latency distributions, retain detailed traces selectively, sample successful paths more aggressively than failures, and run watchdog queries against indexed state and age fields. Telemetry volume, query latency, retention, and cost should be reviewed as part of the design.

## SIGNALS AND DIAGNOSTIC USE OF MTTD, MTTR, AND CUSTOMER IMPACT TIME

Service health engineering relies on core signals such as availability, latency percentiles, throughput, error rates, saturation, dependency health, watchdog results, and customer impact measures [1]-[3]. These signals become more informative when combined with diagnostic metrics:

MTTD describes how long it takes for the team to become aware of a broken approval promise. Detection sources include alerts, watchdogs, synthetic checks, support tickets, and internal engineer observations.

MTTR describes how long it takes to restore the approval promise once a violation is detected. Recovery includes mitigation (feature toggles, rate limiting), repair (fixes to code or configuration), and remediation (reprocessing stranded approvals).

Customer impact time captures the duration during which users are meaningfully affected—for example, the period during which approvals fail to progress or decisions are delayed beyond acceptable bounds.

These metrics should be used diagnostically, not as blunt performance scores. When teams correlate MTTD, MTTR, and customer impact time with detection sources and signals, they can ask targeted questions: Are we discovering broken approval promises through telemetry or only when users call support? Which failure modes are repeatedly detected by synthetic checks, and which are largely invisible until late? Where does recovery rely on manual, fragile procedures?

Diagnostic analysis helps teams prioritize improvements—better watchdogs, more effective alerts, improved runbooks, or resilient design changes—based on how much they reduce real customer impact.

A documented HubSpot incident shows why dependency recovery and workflow recovery need separate measures. During an October 2025 AWS disruption, HubSpot redirected asynchronous tasks including scheduled emails, workflow executions, reports, and data imports and exports to a backup queue. AWS announced service restoration at 3:50 p.m. ET, but HubSpot did not clear the accumulated task backlog until 9:11 p.m., more than five hours later. A replay defect and processing constraints prolonged background-job degradation after the underlying cloud services had recovered [12]. For a service-health review, backlog age, workflow completion, and clearance time remain part of customer impact after dependency availability has been restored.

## RESILIENCY TESTING: DETECT, DEGRADE, RECOVER, ALERT

Resiliency testing validates that systems respond to stress in ways that align with reliability goals. In the context of the approval workflow, teams can design exercises that test four behaviors:

This approach uses established failure-injection and chaos-engineering techniques [11]. The journey-level success criteria include whether the system detects the failure, degrades safely, restores current processing, and completes or explicitly reconciles delayed work.

Detect means that the system and operators become aware of approval promise violations in a timely manner. For example, a backlog of approvals caused by a misbehaving background job should trigger a watchdog alert rather than only appearing in support tickets.

Degrade means that when dependencies slow down or fail, the service degrades in controlled, predictable ways. Approvals might be delayed, but the system should avoid unsafe behavior such as silently dropping documents or giving incorrect statuses.

Recover means that recovery mechanisms such as failover, rollback, and reprocessing restore the approval promise within acceptable time windows. Teams should regularly practice recovery procedures to ensure they work as designed [1], [10].

Alert means that alerts and notifications accurately reflect the situation, reaching the right responders without excessive noise. Reliability engineers can use chaos engineering principles deliberately by introducing controlled faults with a limited blast radius to exercise these behaviors [11].

Resiliency tests might simulate a queue backlog of approval tasks, increased latency or intermittent failure from an external notification provider, stuck background jobs that prevent approvals from advancing, or repeated approval submissions that exercise idempotency. Each test is designed and supervised by humans, with clear rollback plans and success criteria tied to the approval promise.

The goal is not to break systems for its own sake, but to learn how approvals behave under realistic stress and to improve detection, degradation, recovery, and alerting before users experience severe failures.

## WEEKLY SERVICE-HEALTH REVIEWS

The weekly service-health review is a journey-focused operational review. For the document approval system, it brings engineers, SREs, operations, and product stakeholders together to review SLO performance, stalled approvals, incident history, alert behavior, recovery status, and outstanding actions.

A typical agenda includes what happened to approvals over the past week: how many submissions reached final decisions, how the time-to-decision distribution compared to expectations, whether watchdogs detected stranded documents, and which incidents or near misses affected the journey. Teams also examine how alerts behaved—whether they fired when they should, whether too many low-value alerts caused fatigue, and whether any failures went undetected.

Practitioner experience in enterprise SaaS environments suggests that these reviews work best when they stay anchored to two or three key journeys, such as document approval and notification delivery, instead of trying to cover every service. Teams that treat the review as a dashboard tour tend to generate commentary but few changes. Teams that treat it as a reliability planning meeting where each observation is either accepted as "good enough for now" or turned into a backlog item see more sustained improvement.

Crucially, weekly reviews should produce action items, not status theater. The outcome is a set of decisions such as adjusting or adding alerts, refining watchdog thresholds, improving trace propagation in a particular path, updating runbooks based on recent incidents, or planning a resiliency test involving a newly critical dependency. Each action item has an owner and a target date.

Over time, this simple practice turns telemetry into reliability work. Instead of asking whether dashboards look healthy, teams ask whether promises were kept, what failed or degraded, and what they will do about it.

## HUMAN-REVIEWED, AI-ASSISTED HEALTH REPORTING

Preparing weekly service-health reports can be time-consuming. Reliability engineers and SREs sift through metrics, logs, traces, incident records, deployment histories, and support tickets to assemble a coherent picture of what happened. AI can help with this summarization work if used carefully.

In a human-reviewed, AI-assisted workflow, teams provide AI systems with curated evidence: aggregated signals for the approval journey, annotated incident timelines, dependency health summaries, and lists of alerts and watchdog results. AI then drafts summaries, highlights trends, and suggests questions such as "Why did approvals from a particular region show longer decision times last week?"

Inputs should be prepared before they reach the model. Fixed queries compute completion, latency, stuck-state counts, MTTD, MTTR, customer impact time, and week-over-week changes. Incident, deployment, watchdog, and support records are correlated to the review window and relevant workflow or incident identifiers. The model receives this bounded evidence package and drafts the narrative and review questions; it does not query unrestricted production data or calculate authoritative metrics.

Typical failure modes include incorrect trend statements, mixing evidence from different incidents or review windows, treating correlation as root cause, summarizing stale or incomplete data, and exposing sensitive support content. Controls include source references for material claims, timestamps and incident or workflow identifiers, access filtering and redaction before model use, deterministic metric calculation outside the model, and an explicit “insufficient evidence” result when sources conflict.

Human validation is still required. The reviewer checks numerical claims against query output, follows source references for material statements, confirms the review window and incident boundaries, removes unsupported causal language, checks for sensitive content, and approves the proposed actions. That validation is operational work, not a free by-product of automation. Teams should track review time, correction rate, and rejected AI claims to determine whether the pipeline is reducing effort.

## REFERENCE ARCHITECTURE FOR WEEKLY SERVICE-HEALTH REVIEW GENERATION

A weekly service-health review can be generated through a structured pipeline rather than assembled manually from disconnected dashboards. The pipeline begins with review configuration that defines the service, customer journey, owners, review window, signal queries, and watchdog thresholds. A data collection stage gathers telemetry, workflow state, watchdog results, incident records, deployment history, support tickets, and reliability backlog items. A normalization and correlation stage then maps raw evidence to customer journeys, service promises, detection sources, time windows, and affected workflow states.

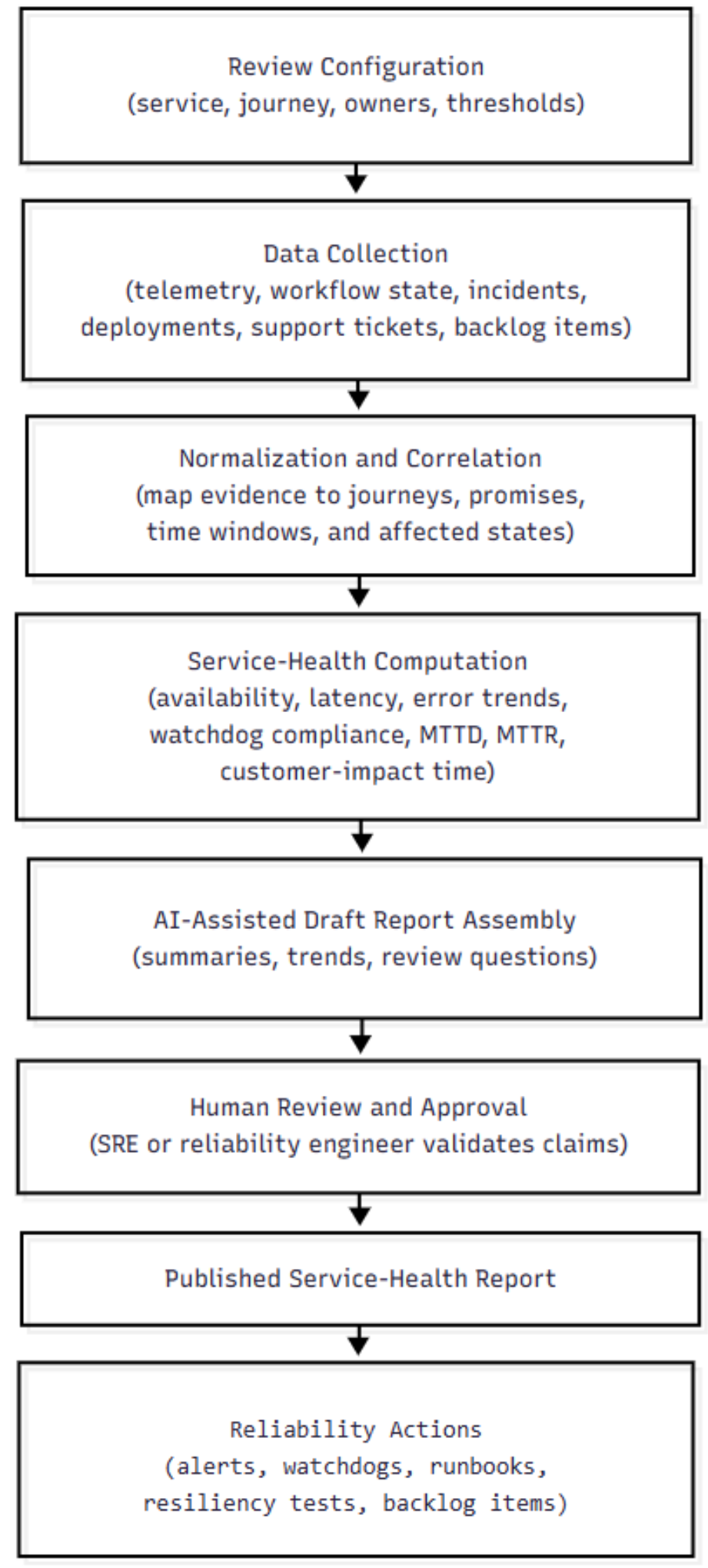

**Figure 1.** Reference architecture for human-reviewed, AI-assisted weekly service-health review generation.

The end-to-end reporting pipeline is shown in Fig. 1. The pipeline converts journey-level telemetry, workflow state, incident records, deployment history, support tickets, and backlog items into a human-reviewed service-health report and reliability actions.

The normalized evidence feeds a service-health computation stage that summarizes availability, latency, throughput, error-rate trends, stuck workflow counts, watchdog results, MTTD, MTTR, customer impact time, and week-over-week changes. A report assembly stage organizes these findings into a draft review containing an executive summary, journey-health summary, incident and near-miss summary, watchdog findings, reliability risks, and proposed action items. AI may assist with summarization and review questions, but engineers validate every claim before publication. The final review is published to an internal portal or wiki, and its action items feed back into alerts, watchdogs, runbooks, resiliency tests, and design improvements.

## LESSONS FROM ENTERPRISE SAAS PRACTICE

Practitioner experience with large SaaS-style workflows suggests several recurring lessons.

One lesson is that journey ownership matters more than component ownership. When a document approval journey crosses teams, regions, and services, someone still needs to be explicitly accountable for whether approvals complete as promised. Without that, gaps such as notifications never firing for a subset of tenants can persist because each team's local metrics appear healthy.

A second lesson is that silent failures in asynchronous paths are often under-detected. Queues drain, workers run, and infrastructure appears healthy, yet approvals sit in "pending" or "in review" indefinitely. The teams that make progress here treat watchdogs as first-class signals and give them clear thresholds and escalation paths.

A third lesson is that reliability work must compete fairly with feature work. Service-health reviews that only describe problems, without tying them to explicit backlog items and trade-offs, tend to be ignored. Reviews that lead to a small number of prioritized tasks such as adding a watchdog, tightening an alert, or practicing a recovery drill fit more naturally into planning cycles.

These patterns reflect common challenges in operating multi-tenant, multi-region workflows at scale.

## STARTING SMALL

Service health engineering can look ambitious when described end-to-end. In practice, teams can start with one critical workflow.

One pragmatic path is to choose a single workflow such as document approval, write down a simple promise in user terms, and identify three to five signals that best indicate whether that promise is being kept. From there, teams can add one watchdog for stranded approvals, run one small resiliency test that exercises detect–degrade–recover–alert behavior for a single dependency, and introduce one short weekly review focused only on this journey.

On the reporting side, teams can begin with a manual health summary based on curated telemetry and incident records, then gradually introduce AI assistance to help assemble and highlight the data they already use. The key is to keep humans in charge of interpretation and decisions.

As confidence grows, the same patterns can be extended to other workflows—notification pipelines, billing runs, or background reconciliation jobs without changing the core idea: define promises, connect telemetry to those promises, and schedule reliability work around what users actually experience.

## FROM TELEMETRY TO RELIABILITY PRACTICE

Service Health Engineering brings established reliability practices together around end-to-end journey completion. It does not replace SLOs, operational reviews, or chaos engineering. It applies them to silent failures, stranded work, and incomplete recovery in asynchronous systems.

In the document approval example, component telemetry helps explain system behavior, while completion, time-to-decision, watchdog results,

MTTD, MTTR, and customer impact time show whether the service promise was met and recovery was complete. Weekly reviews and resiliency tests turn this evidence into owned engineering actions.

AI can help assemble and summarize this evidence, but service-health metrics should be calculated deterministically, and engineers must validate sources, incident boundaries, causal claims, and sensitive content. A practical starting point is one critical journey, a small set of SLIs and SLOs, one workflow watchdog, and a recurring operational review.

## ⬛ REFERENCES

**Siva Rama Krishna Varma Bayyavarapu** is currently with Docusign Inc., Indianapolis, IN, USA, where he works on distributed enterprise systems, cloud platforms, and service reliability engineering. He received the M.Tech. degree from Jawaharlal Nehru Technological University Anantapur, India. His research interests include service health engineering, distributed systems, incident detection and response, enterprise system integrations, cloud-native architectures, and AI-assisted software operations. He has authored and reviewed technical work in areas related to software reliability, enterprise systems, artificial intelligence, and distributed computing, and has served as a reviewer for IEEE and other technical venues. He is an IEEE Senior Member and is actively involved in IEEE technical and professional activities. Contact him at siva.bayyavarapu@ieee.org.